\documentstyle[12pt]{article}
\begin{document}
\baselineskip=24pt
\date{\ }
\title{Excitonic Phase Transition in Electronic Systems}
\author{Mucio A.Continentino \\
Instituto de F{\'{\i}}sica, Universidade Federal Fluminense \\
Campus da Praia Vermelha, Niter\'oi,  24.210-340, RJ, Brazil \\
and \\
National High Magnetic Field Laboratory \\
Florida State University, 1800 E.Paul Dirac Dr. \\
Tallahasse, Fl 32306, USA \\
\\Gloria M. Japiass\'u \\
Instituto de F{\'{\i}}sica, Universidade Federal do Rio de Janeiro \\
Caixa Postal 68.528, Rio de Janeiro, 21.945-970, RJ, Brazil \\
\\Am\'os Troper \\
Centro Brasileiro de Pesquisas F{\'{\i}}sicas \\
Rua Dr. Xavier Sigaud 150, \\
Rio de Janeiro, 22.290-180, RJ, Brazil}
\maketitle
\baselineskip=24pt
\begin{abstract}
Although predicted theoretically excitonic phase
transitions have never been observed.
Recently it has been claimed that they can occur in some strongly
correlated materials.
We  study  the possibility
of an excitonic transition in a  two-band model and
show that a true phase transition never occurs in the
presence of hybridization.
We suggest an alternative
interpretation for these experiments based on the opening or
closing
of a hybridization gap at a critical pressure.
\end{abstract}
\newpage

Recently Wachter and colaborators \cite{Wachter,wachter2} have claimed to
observe a transition to an excitonic insulator in a doped
narrow-band-gap semiconductor at moderate pressures and in
a strongly correlated metal \cite{wachter2}.
The evidence for this transition is obtained by monitoring
the resistivity close to the metal-insulator
transition as a band-gap is
continuously closed or opened by external pressure. The resistivity
at low temperatures showed a huge peak as a function of applied
pressure which they attibuted to the long predicted
excitonic phase transition \cite{Mott}. They also presented
data of the Hall constant which reveals that this resistivity
anomaly is caused by a reduction in the number of carriers.

With renewed interest in this subject  motivated by these
challenging experiments \cite{Wachter,wachter2} and
the close connection of the materials
which have been investigated with Kondo insulators \cite{aeppli}, we
study the possibility of this excitonic transition in presence of
hybridization in a two-band model. We point out that no sharp
excitonic phase transition occurs when external pressure is applied.
The reason is that the hybridization which depends on pressure
acts as a conjugate field to the order parameter of the excitonic
phase and destroys this transition \cite{Continentino}. The effect
is similar to that of a ferromagnetic system in the presence of an
uniform external magnetic field.
The mixing term however is strongly renormalized
by excitonic correlations. We
suggest an  interpretation for
the experiments of Wachter et al.  on $Sm_{0.75}La_{0.25}S$
\cite{wachter2} on the basis of a metal-insulator  (MI) transition
associated with the appearance of a hybridization gap at a critical pressure.
This metal-insulator  transition
is in  the universality class
of the  {\em density-driven transitions}, for which
the relevant critical exponents  have been obtained previously
\cite{Continentino}.
This  seems to hold  even in the presence of correlations,  as we find here,
as long as the MI transition  is not accompanied by the appearance of long
range
magnetic order  \cite{Continentino}.
A characteristic feature of this  transition
is that the gap opens or closes
linearly with pressure near the critical pressure. This is a
direct consequence of the fact that
the gap exponent \cite{Continentino}    assumes the value $\nu z = 1$ .

The Hamiltonian which describes  our system is:
\begin{equation}
H = \sum_{k} \epsilon_{k}^a a_{k}^\dagger a_{k} +
\sum_{k} \epsilon_{k}^b b_{k}^\dagger b_{k} +
\sum_{k} V_k (a_{k}^\dagger b_{k} + b_{k}^\dagger a_{k}) -
\sum_{k,k \prime , q} G (q) a_{k+q}^\dagger a_{k} b_{k \prime-q}^\dagger
b_{k \prime}
\end{equation}
where $\epsilon_{k}^a$ and $\epsilon_{k}^b$ represent the energies
for electrons   in the narrow $a-band$
and in the large conduction $b-band$,  respectively.
The operators $a_{k}^\dagger$, $a_{k}$ create and destroy
electrons in the narrow band and $b_{k}^\dagger$, $b_{k}$
are creation and annihilation operators of electrons in the wide conduction
band.  V is the mixing term, which arises from the
crystalline potential  and  G is the effective attractive
interaction between spinless
electrons and holes \cite{falikov}.
For the situation we are interested this is the most important
interaction. Note that we should have taken into account in Eq.1, the Coulomb
repulsion
between the electrons in the narrow a-band. Since, however, as we argued before
\cite{Continentino} these interactions are irrelevant, in the renormalization
group sense, for the metal-insulator transition we are going to study we do
not consider it for simplicity. In fact within the approximation which is
generally used
to treat the Hamiltonian above \cite{Mott} and that we also use in this Letter,
these interactions merely shift the energy of the {\em a-electrons \/}.

We point out that we adopt here a different approach to the excitonic
problem \cite{Mott}. Instead of starting with two hybridized bands and
the electron-hole
attraction \cite{Mott}, we include explicitly the hybridization in the starting
Hamiltonian
to imply that it should be treated together with the many-body correlations
and not separetely. The main difference is that within our approach
the gap which eventually opens or closes  as will be shown below and is {\em
experimentally
acessible \/} turns out to be a
truly many-body quantity. In the standard treatment \cite{Mott} where the bands
are diagonalized separately from the correlations one arrives at the
artificial situation where the gap is a simple one-body effect.

If it were not for the attractive term, the
Hamiltonian given  by Eq.1  could be exactly diagonalized giving rise to two
hybrid bands \cite{Japi}. However the many-body term due to the effective
attractive interaction $G$ makes this a difficult
problem for which an approximation must be introduced.
We shall employ the Green's function method \cite{green} to obtain
the order parameter associated with the excitonic phase, namely,
$ \Delta = \sum_k \langle b_{k}^\dagger a_{k} \rangle $.
When calculating the equation of motion for the Green's function
$ \langle \langle a_{k};b_{k}^{\dagger}  \rangle \rangle_\omega$ we
find it
gives rise to new propagators.  Introducing a
a convenient mean-field approximation \cite{Mott}:
$$\langle\langle a_{k-q} b_{k \prime-q}^\dagger b_{k\prime};b_{k}
^\dagger \rangle\rangle_\omega \approx
\langle a_{k-q} b_{k \prime -q}^\dagger \rangle \langle\langle
b_{k \prime};b_{k}^\dagger \rangle\rangle_\omega $$
\begin{equation}
\langle \langle a_{k \prime + q}^\dagger a_{k \prime} b_{k \prime +q} ;
b_k^\dagger \rangle \rangle_\omega \approx
- \langle a_{k \prime +q}^\dagger b_{k+q}  \rangle \langle \langle a_{k \prime}
;
b_{k} \rangle \rangle_\omega .
\end{equation}
we obtain a closed set of equations which can be solved to yield
\begin{equation}
\langle\langle a_{k};b_{k}^\dagger \rangle\rangle_\omega =
\frac{\tilde{V}_k}{[ ( \omega
-\epsilon_{k}^a ) ( \omega-\epsilon_{k}^b)
- \tilde{V}_k ^2 ]}
\end{equation}
where $ \tilde{V}_k = V + G \Delta_k $, with $\Delta_{k} =
\langle b_{k}^\dagger a_{k} \rangle$.
We have neglected the k-dependence of $G$ and $V$ for simplicity.
The new energies of excitation of the system are given by the poles
of the  above Green's function,  i.e., by  the roots of the equation
\begin{equation}
( \omega - \epsilon_{k}^a)( \omega - \epsilon_{k}^b) -
{ \tilde{V}_k}^2 = 0
\end{equation}
so that
\begin{equation}
\omega_{1,2}(k) = \frac{1}{2} \{ \epsilon_{k}^a +
\epsilon_{k}^b \pm \sqrt{( \epsilon_{k}^a - \epsilon_{k}^b )^2 +
4 {\tilde{V}_k}^2 } \}
\end{equation}
The excitonic propagator can be rewritten as:
\begin{equation}
\langle\langle a_{k};b_{k}^\dagger \rangle\rangle_\omega =
\frac{\tilde{V}_k} {\omega_1(k) - \omega_2(k)}
\{ \frac{1}{ \omega - \omega_1(k) } - \frac{1} { \omega - \omega_2(k)}\}
\end{equation}
from which we obtain
\begin{equation}
\Delta_k =
\frac{\tilde{V}_k}{ \vert { \omega_1(k) - \omega_2(k)} \vert}
\int {d \omega} {f(\omega)} \left\{ \delta[ \omega - \omega_1(k)] -
\delta[ \omega - \omega_2(k)] \right\}.
\end{equation}
where $f(\omega)$ is the Fermi function.

In order to obtain explicit results for the excitonic order parameter
we adopt the homothetic band model \cite{kishore} which consists
in taking
\[
\left\{ \begin{array}{ll}
        \epsilon_k^b = \epsilon_k \\ \epsilon_{k \sigma}^a =
        \alpha \epsilon_k + \beta
        \end{array}
        \right. \]
The quantity $\alpha$ $( \alpha < 1)$ may be interpreted as taking
into account the different effective masses of the electrons in the
narrow $a-$ band and the large $b-$ band, i.e. $(m_b/m_a) = \alpha$.
The quantity
$\beta$ gives the shift of the narrow band with respect to
the large band.

Now we introduce two new functions
$g_1(\omega)$ and $g_2(\omega)$ \cite{Japi} through the following equation
\begin{equation}
[\omega- \omega_1(k)][\omega - \omega_2(k)] =
\alpha [g_1(\omega) - \epsilon_k][g_2(\omega) - \epsilon_k]
\end{equation}
from which we get
\begin{equation}
g_{2,1}( \omega) =
\frac{1}{2 \alpha} \{ (1 + \alpha) \omega -\beta
\pm \sqrt{ [(\alpha-1)\omega + \beta]^2 + 4 \alpha \tilde{V}^2} \}
\end{equation}
The energies of the bottom of the hybrid bands correspond to
$g_i(E_{B}^i)=0$ with $i=1,2$
\begin{equation}
E_{B}^{2,1} = \frac{1}{2} \{ \beta \pm {[ \beta^2 + 4 \tilde{V}^2]}
^{\frac{1}{2}} \}
\end{equation}
while the energies of the tops are obtained when $g_i(E_{T}^i)=D$, where
$D$ is the bandwidth of the large conduction $b-$ band,
\begin{equation}
E_{T}^{2,1} = \frac{D}{2} \{ (1+ \alpha) + \frac{\beta}{D} \pm
[(( \alpha - 1) + \frac{\beta}{D} )^{2} + 4 ( \frac{\tilde{V}}{D} )^2 ]
^{\frac{1}{2}} \}
\end{equation}
For $\tilde{V} = 0$, we have $E_B^1 = 0$, $E_T^1 = D$, $E_B^2 = \beta$ and
$ E_T^2 = \alpha D + \beta$, which are in agreement with the definitions
of $ \alpha$ and $ \beta$. Considering the new functions $g_i (\omega)$ we
obtain the
following equation for the excitonic order parameter
%\begin{equation}
$$\Delta = \frac{1}{ \alpha}  \int{  d\omega { \frac{\tilde{V}}
{|g_{1}(\omega) - g_{2} (\omega)|} } f(\omega) \left\{ N [g_{1}(\omega)]  -
N [g_{2}(\omega)] \right\}}$$
%\end{equation}
where
$N(\omega)=\sum_k \delta(\omega - \epsilon_k)$ and
$$\vert{ g_1(\omega) - g_2(\omega)} \vert = \frac{1}{\alpha}
[(\omega (1- \alpha) - \beta)^2 + 4\alpha \tilde{V}^2]
^\frac{1}{2}.$$

We also obtain an expression for the gap $ \Delta_G $ between the two
bands as a function of the hybridization $V$ and the electron-hole
interaction $G$. It
corresponds to  the difference in energy between the top of
the first hybrid band  $E_T^1$ and the
bottom of the second $E_B^2$ :
\begin{equation}
\frac{\Delta_G}{D} = \frac{1}{2} \{ [( \alpha -1 + \frac{\beta}{D})^2 +
4 (\frac{\tilde{V}}{D})^2]^{\frac{1}{2}} +
[ ( \frac{\beta}{D} )^2 + 4 (\frac{\tilde{V}}{D} )^2]^{\frac{1}{2}} - (1+
\alpha) \}.
\end{equation}
Consequently for a two-band system the opening of a hybridization gap,
contrary to what occurs for the Anderson lattice model \cite{varma}, requires a
critical
value of renormalized hybridization  $\left( \frac{\tilde{V}}{D} \right)_c$
   given by
\begin{equation}
\left( \frac{\tilde{V}}{D} \right)_c = \frac{1}{2} \{ \frac{[2 \alpha -
\frac{\beta}{D}( \alpha -1)]
^2}{(1+ \alpha)^2} - ( \frac{\beta}{D} )^2 \}^{\frac{1}{2}}.
\end{equation}
Notice that for $\alpha \rightarrow 0$, $\tilde{V}_c \rightarrow 0$ as expected
for a
collection of localized levels.
In this case also $\Delta_G \propto {\tilde{V}}^2$ \cite{varma}
contrary to our two-band problem where
close to $\left( \frac{\tilde{V}}{D} \right)_c$
we find
$$\frac{\Delta_G}{D} = \left| \left( \frac{\tilde{V}}{D} \right) -
\left( \frac{\tilde{V}}{D} \right)_c \right|$$
so that the
gap opens or closes
linearly near $\tilde{V}_c$, i.e. $\nu z = 1$, as in the non-interacting case
\cite{Continentino}.
Within the assumption that
$\vert \left( \frac{\tilde{V}}{D} \right) -
\left( \frac{\tilde{V}}{D} \right)_c \vert \propto \vert
P - P_c \vert$ where $P_c$ is the critical pressure, this result, i.e.,
$\Delta_G \propto
\vert P - P_c \vert$, describes the observed
behavior for $Sm_{0.85}La_{0.15}S$ \cite{wachter2}.
We emphasize that
the relevant variable here is $\left( \frac{\tilde{V}}{D} \right)$.
This can  increase or decrease with pressure, for a given pressure range, so
that a gap can either open or close
depending on the relative pressure dependence of
the hybridization and bandwidth.
So our approach may be also useful to describe the metal-insulator
transition in  the compound $SmB_6$
\cite{brandt} \cite{cool}.

In order to study the possibility of an excitonic transition in the presence of
hybridization we  investigate the case of two square symmetric bands
with respect to the Fermi level fixed
at $\mu=\frac{3}{4}D$, such that $\alpha=1$ and $\beta=\frac{D}{2}$.
This corresponds to a divalent
semi-metal with two electrons per site, where $\mu$ is at the crossing of
the bands \cite{Continentino}.
In this case, we get a simple expression for the $T=0$  excitonic
order parameter
\begin{equation}
\Delta = \frac{V}{G - G_c}.
\end{equation}
Then there is a critical value for the electron-hole  attraction $G_c$,
$G_c = D$, for which the excitonic order parameter is different from
zero  even in the absence of hybridization. This is formally similar to a
Stoner-like
criterion for the appearance of ferromagnetic order in a
metallic system.  In the present case, the hybridization
plays the role of the uniform
magnetic field conjugate to the order parameter $ \Delta $.
The fact that a k-independent, local  hybridization $V$
acts as the conjugate field of the excitonic order parameter,
independently of
any particular approximation, can be directly seen
from the Hamiltonian, Eq. 1, with  the one-body
mixing term   written as
$V \sum_{k} (a_{k}^\dagger b_{k} + b_{k}^\dagger a_{k}) = V \Delta$.
This shows that $V$ couples directly to $\Delta$.
For the case of $d-f$ hybridization there are different mechanisms, for
example phonons, which can provide local mixing between these orbitals
with different symmetries \cite{nunes}.
Consequently we do not expect a sharp phase transition to an excitonic phase to
occur  whenever an external parameter
which changes the hybridization, as external pressure for example  is varied.
There is no singularity in any physical quantity
when $G$ approaches  $G_c$, whenever $V \neq 0$,
contrary to what the simple mean-field equation above suggests.
However we expect
strong renormalization of the physical
quantities, as  the hybridization, due to excitonic correlations for $G \approx
G_c$.
The only zero temperature phase transition which can occur in
our model  as pressure is varied  is a metal-insulator
transition associated with the opening of a hybridization gap
at a critical value of the ratio $(V/D)_c$.
For the divalent semi-metal discussed above this occurs at
$(V/D)_c = \left[(\surd{3}/4) - (G/D)\Delta \right]$,
%$\left( \frac{V}{D} \right)_c = {\frac{ \surd {3}}{4}} - \frac{G}{D} \Delta$
%\end{equation}
where due to the excitonic correlations this value is
renormalized with respect to that of the non-interacting system
\cite{Continentino}.

For completeness we give  the self-consistency equation for
the renormalized hybridization for general $\alpha$ and $\beta$ and
square bands
\begin{equation}
V = \tilde{V} \left[ 1 - \frac{G}{D(1 - \alpha)}Ln \vert \frac{\cal A}{\cal B}
\vert \right]
\end{equation}
with
$${\cal A( \alpha, \beta)} =  E^2_B (1 - \alpha) - \beta +
{ \left\{ {\left[ E^2_B (1 - \alpha) -
\beta \right] }^2
+ 4 \alpha {\tilde{V}}^2 \right\} }^{\frac{1}{2}}$$
and
$${\cal B( \alpha, \beta)} =  E^1_B (1 - \alpha) - \beta + { \left\{ {\left[
E^1_B (1 - \alpha) - \beta \right] }^2
+ 4 \alpha {\tilde{V}}^2 \right\} }^{\frac{1}{2}}$$
with $E^1_B$ and $E^2_B$ given before.
We notice that this is independent
of $ \mu $ and for $ \beta = D/2 $ and in
the limit $\alpha \rightarrow  1$ it
gives correctly the previous equation for $\Delta$ in the symmetric,
divalent semi-metal case.
In the limit $\alpha \rightarrow 0$, it reduces to the expression for
the renormalized hybridization in the Anderson lattice, i.e., for
a collection of local levels \cite{nunes} \cite{leder}.

We give now the scaling results for the
properties of the Fermi liquid in the metallic phase close to the
metal-insulator transition. We find that the thermal mass
$m_T$, defined as the coefficient of the linear term of the specific heat
scales
as $m_T \propto {\vert P - P_c \vert}^{\frac{d}{2} - 1}$ and in three
dimensions
vanishes as $m_T \propto {\vert P -P_c \vert}^{1/2}$.
The same scaling is found for the uniform
susceptibility $\chi_0$, the compressibility $\kappa$ and
the density of states at the chemical potential.
The number of carriers $n_c \propto {\vert P - P_c \vert}^{\frac{d}{2}}$.
The behavior of the thermal
mass obtained above  is  opposite to that found in
heavy fermions \cite{cont2}, where it  is enhanced
as the system approaches the critical point.
The characteristic or coherence temperature in the metallic phase scales as
$T_c \propto (P - P_c)$ since $\nu z =1$.
The existence of a small coherence temperature close
to the density-driven phase transition gives rise in
the presence of electron-electron interactions to a
significant $T^2$ term in the resistivity even
in wide band materials as $Yb$ \cite{jullien}.
If we write $\rho \approx
\rho_0 +  A T^2$ for $T << T_c$, then the
coefficient $A$ scales as $A \propto {T_c}^{-2} \propto {(P - P_c)}^{-2}$.

At the critical pressure, $P = P_c$, we  find
non-Fermi liquid behavior  with the
specific heat vanishing with temperature as $C \propto T^{3/2}$ and
$\chi_0 \propto T^{1/2}$. In actual systems, states in the
tails of  the density of states due
to impurities or  disorder and which give rise to a saturation of
the resistivity at the lowest temperatures may spoil this simple behavior.

In conclusion the possibility of an excitonic phase transition  in a
two-band model has been investigated.
The dominant interaction was taken to be the electron-hole attraction.
We  argued that a phase transition to an
excitonic phase never occurs in the presence of hybridization
since the one-body mixing term
acts as a conjugate field to the order parameter of this phase.
We  suggested an alternative interpretation of the experiments of Wachter et
al.
on $Sm_{0.75}La_{0.25}S$.
Our  approach gives rise to a gap that varies linearly close to the critical
pressure, i.e., $\Delta_G \propto (P - P_c)$ as observed
\cite{wachter2,brandt}.
We have
derived the properties of the Fermi liquid in the metallic phase close to the
transition
and predicted non-Fermi liquid behavior at the critical pressure.  \\
\\
{\bf Acknowledgments:}
We thank Conselho Nacional de Desenvolvimento
Cien\-t{\'{\i}}\-fi\-co e Tecnol\'ogico - CNPq - Brazil, for partial
financial support.


\newpage

\begin{thebibliography}{99}

\bibitem{Wachter} J. Neuenschwander and P. Wachter, Phys. Rev. B {\bf 41},
12693
(1990); B. Bucher, P. Steiner, and P. Wachter, Phys. Rev. Lett. {\bf 67}, 2717
(1991); P.Wachter and A.Jung, IEEE Transactions on Magnetics {\bf 30}, 954
(1994).

\bibitem{wachter2} P.Wachter, A.Jung and P.Steiner, Phys.Rev. {\bf B51}, 5542
(1995).

\bibitem{Mott} N.F.Mott, Phil. Mag. {\bf 6}, 287 (1961); D.Jerome, T.M.Rice
and W.Kohn, Phys.Rev. {\bf 158 \/}, 462 (1967); B.I.Halperin and
T.M.Rice, Rev. Mod. Phys. {\bf 40}, 755 (1968);

\bibitem{aeppli} G.Aeppli and Z.Fisk, Comments Cond. Mat. Phys. {\bf 16}, 155
(1992).

\bibitem{Continentino} M.A. Continentino, Phys. Lett.  {\bf A197}, 417 (1995).


\bibitem{falikov} R.Ramirez, L.M.Falicov, and J.C.Kimball, Phys. Rev. B {\bf
2}, 3383 (1970).

\bibitem{Japi} M.A. Continentino,
G.M. Japiass\'u, and A. Troper, J. Appl. Phys. {\bf 75}, 6734 (1994);  M.A.
Continentino,
G.M. Japiass\'u, and A. Troper,
Phys. Rev. B {\bf 49}, 4432 (1994).


\bibitem{green} See, for example, A.L.Fetter and J.D.Walecka, {\em Quantum
Theory of Many-Particle Systems } McGraw-Hill, Inc. 1975 USA.

\bibitem{kishore} R. Kishore and S.K. Joshi, Phys. Rev. B {\bf 2}, 1411 (1970).

\bibitem{varma} C.M.Varma and Y.Yafet, Phys.Rev. {B19}, 2950 (1975).

\bibitem{brandt} I.V.Berman et al., JETP Lett. {\bf 38}, 477 (1983); J.Beille
et
al., Phys.Rev.{\bf B28}, 7397 (1983).

\bibitem{cool} Very recently Cooley et al., Phys.Rev.Lett. {\bf 74}, 1629
(1995), through a
different analysis of the resistivity curves  concluded that the
metal-insulator
transition as a function of pressure in $SmB_6$ is discontinuous.

\bibitem{nunes} M.D. Nunez-Regueiro and M. Avignon, J. Magn. Magn. Mater.
{\bf 47-48}, 302 (1985); M.D.Nunez-Regueiro, PhD Thesis, {\em Universite
Scientifique et
Medicale de Grenoble}, 1985.

\bibitem{leder} H.J.Leder, Sol.St.Comm., {\bf 27}, 579 (1978).

\bibitem{cont2} M.A.Continentino, Phys.Rep. {\bf 239}, 179 (1994).

\bibitem{jullien} R.Jullien and D.Jerome, J.Phys.Chem.Solids, {\bf 32}, 257
(1971).

\end{thebibliography}
\end{document}